\documentclass[conference]{IEEEtran}
\usepackage{graphicx}
\usepackage{amsmath, amssymb}
\usepackage{url}
\usepackage{hyperref}
\usepackage{mciteplus} % Handles multiple citations in a single box
\usepackage{tikz}
\usetikzlibrary{positioning,arrows.meta,fit}
\usetikzlibrary{calc}
\usepackage{listings}
\usepackage{xcolor}
\usepackage{cite}
\usepackage{float}
\usepackage{placeins}
\definecolor{codegray}{rgb}{0.5,0.5,0.5}
\definecolor{codepurple}{rgb}{0.58,0,0.82}
\definecolor{backcolour}{rgb}{0.95,0.95,0.92}

\lstdefinestyle{mypython}{
    backgroundcolor=\color{backcolour},
    commentstyle=\color{codegray},
    keywordstyle=\color{blue},
    numberstyle=\tiny\color{codegray},
    stringstyle=\color{codepurple},
    basicstyle=\ttfamily\footnotesize,
    breaklines=true,
    captionpos=b,
    numbers=left,
    numbersep=5pt,
    showspaces=false,
    showstringspaces=false,
    tabsize=4,
    language=Python
}

\begin{document}

% --- Title and Authors ---
\title{Quantum Encryption Resilience Score (QERS) for Computer System Processes, IoT, and IIoT Devices Across MQTT, HTTP, and HTTPS}

\author{
    \IEEEauthorblockN{Jonatan Rassekhnia}
    \IEEEauthorblockA{
        Luleå University of Technology (LTU) \\
        Department of Computer Science, Electrical and Space Engineering \\
        Sweden \\
        Email: rasjon-0@@student.ltu.se
    }
}

\maketitle

% --- Abstract ---
\begin{abstract}
This research presents an implementation of the Quantum Encryption Resilience Score (QERS) for computer processes/systems and IoT/IIoT environments across the communication protocols MQTT, HTTP, and HTTPS using constrained devices (ESP32-C6 and an ARM-based 64-bit Raspberry Pi CM4). The devices communicate using a client--server model, in which QERS is computed through three proposed formulas applied to experimental measurements derived from both performance and security metrics. These metrics include CPU utilization, RSSI, TLS handshake latency, energy consumption, and encryption key size. By unifying these factors into a single comparative score, QERS enables systematic evaluation across communication protocols and post-quantum cryptographic schemes. This research is complementary to the initial QERS formulation and contributes additional empirical validation, providing insight into secure and efficient deployment of constrained computing, IoT, and industrial IoT systems in post-quantum-enabled environments.
\end{abstract}


% --- Introduction ---
\section{Introduction}
The rise of quantum computing threatens classical cryptographic schemes such as RSA and ECC, which underpins most current communication security in computers, IoT, and IIoT systems \cite{sen2025securityprivacymanagementiot}. As a result, Post-Quantum Cryptography (PQC) has been proposed to secure communications against quantum attacks \cite{kappler2022post, Joseph2022}. Computer processes/systems, IoT, and IIoT devices are typically resource-constrained with limited CPU, memory, and energy budgets, making the performance-security trade-off critical when deploying PQC \cite{bankar2025next,11059522}. Protocols such as MQTT, HTTP, and HTTPS are widely used in computer systems, IoT, and IIoT globally, but their behavior under PQC-enabled cryptography remains insufficiently evaluated \cite{10988807,rahman2025comprehensive}. 
\vspace{0.1 cm}

Measuring latency, throughput, CPU/memory usage, TLS handshake times, and energy consumption in real devices provides essential insights for selecting optimal protocols under PQC \cite{10382535}. Despite existing studies on PQC performance, there is no standardized metric to quantify the combined impact of PQC on protocol performance, energy, and security in traditional computation processes/systems, IoT and IIoT networks \cite{ulanov2024multi,11132566}.
\vspace{0.1 cm}

To address this gap, the Quantum Encryption Resilience Score (QERS) is proposed, a unified metric that integrates latency, CPU and memory usage, energy consumption, TLS handshake cost, and encryption key size, enabling systematic comparison across communication protocols and constrained IoT/IIoT devices \cite{lopez2025evaluatingpostquantumcryptographicalgorithms, KOTANGALE2025103643}.
\vspace{0.1 cm}

\subsection{Research Question:} How can the Quantum Encryption Resilience Score (QERS) be used to quantify and compare the performance and security trade-offs of MQTT, HTTP, and HTTPS on constrained Computer Processes/Systems, IoT, and IIoT devices under post-quantum cryptography?
\vspace{0.1 cm}

\subsection{Contributions}
The main contributions of this paper are as follows:
\begin{enumerate}
    \item A systematic experimental evaluation of MQTT, HTTP, and HTTPS on resource-constrained IoT devices under Post-Quantum Cryptography (PQC),
    measuring latency, CPU usage, memory overhead, TLS handshake cost, and  energy consumption across multiple operating conditions.
    \item The formulation and validation of the \textit{Quantum Encryption Resilience Score (QERS)}, a unified multi-metric scoring framework that integrates performance, reliability, and cryptographic strength into a single normalized indicator for PQC-enabled communication systems.
    \item A reproducible end-to-end measurement methodology combining embedded devices, protocol instrumentation, and data-driven analysis, enabling objective comparison of PQC readiness across IoT protocols and deployment scenarios.
\end{enumerate}
\vspace{0.1 cm}

% --- Related Work ---
\section{Related Work}
This section reviews prior research on IoT/IIoT communication protocols with post quantum cryptography (PQC) performance, highlighting gaps motivating the Quantum Encryption Resilience Score (QERS).
\vspace{0.1 cm}

\subsection{MQTT Performance and Security Studies}
MQTT is widely adopted for lightweight IoT communication due to its low bandwidth and QoS-based reliability features \cite{app11114879, s22228852}. Previous research measured MQTT latency, jitter, and packet loss under varying QoS levels, showing that MQTT performs efficiently in low-power and contained environments \cite{8914552,8971097}. Energy consumption of MQTT clients on microcontrollers has been characterized for different message rates and payload sizes, revealing trade-offs between throughput and battery usage \cite{8765692}. Limited research has investigated MQTT under PQC, primarily focusing on memory overhead introduced by KEM/SIG key operations \cite{jcp3030021}. Overall, these studies lack a unified multi-metric evaluation, making cross-protocol comparisons challenging \cite{Kumar2025}.
\vspace{0.1 cm}

\subsection{HTTP Performance and Security Studies}
HTTP is commonly used in IoT for REST APIs and dashboard communication, but its performance is impacted by payload and header overhead \cite{7184865, 8620130}. Benchmarking studies have evaluated HTTP latency and CPU usage in constrained devices, showing significant performance degradation under frequent requests or large payloads \cite{8400067}. Certain research has extended these analyses to PQC-enabled HTTP, reporting increased TLS handshake times and key-processing overhead for secure HTTP connections \cite{10.1145/3465481.3465747}. Energy efficiency studies indicate HTTP is more resource-intensive than MQTT due to frequent connection setup and lack of persistent sessions in classical implementations \cite{CAIAZZA2024101871, s25196042}. However, no prior study combines these metrics into a normalized, unified score for PQC-readiness evaluation \cite{Silva16032021}.
\vspace{0.1 cm}

\subsection{HTTPS Performance and Cryptographic Overhead Studies}
HTTPS introduces TLS security, providing confidentiality and integrity, but with substantial cryptographic overhead, especially under PQC key exchanges \cite{10.1007/978-3-030-44223-1_5, 10.1145/3465481.3465747}. TLS-1.3 PQC implementations have been bench-marked on ARM and ESP32 devices, revealing increases in handshake latency, CPU usage, memory footprint, and energy consumption \cite{11008381, 10756206}. Studies measuring HTTPS performance under PQC report that message throughput and latency are heavily affected by large KEM/SIG keys, especially in constrained devices \cite{10821264}. Despite these insights, research still lacks a systematic framework to compare HTTPS performance with MQTT and HTTP under PQC across multiple metrics \cite{8597401}. 
\vspace{0.1 cm}

\subsection{Security and Resilience Metrics in IoT}
IoT security scoring frameworks often include latency, packet reliability, and key management considerations \cite{11197543, Lezzi2025}. However, they generally do not integrate energy consumption, RSSI, or CPU usage into a single normalized metric suitable for PQC evaluation \cite{10.1145/3587135.3592821}. Benchmark research exists for PQC algorithms on constrained devices, but they usually evaluate individual metrics in isolation, without a multi-criteria decision-making approach \cite{11136103}.
\vspace{0.1 cm}

\subsection{Machine Learning for Multi-Metric Security Evaluation}

Related research has shown that machine learning can support the evaluation of complex security and performance trade-offs in IoT and
networked systems \cite{al2020survey, hussain2020machine, kataria2022ai}.Traditional analytical models struggle to capture nonlinear relationships between latency, cryptographic overhead, energy consumption, and wireless conditions, especially when post-quantum
cryptography is deployed \cite{gencay1997nonlinear, darzi2024pqc}. Supervised learning methods such as Random Forests have been widely used to predict network performance, energy usage, and protocol behavior from noisy measurement data \cite{huo2021performance, tekin2023energy}.Machine learning has also been applied to estimate cryptographic and TLS-related overhead when protocol parameters or key sizes change \cite{de2018identifying, dowling2015cryptographic}. In multi-criteria decision systems, learning-based weighting and regression models can complement MCDA by improving stability and trend estimation under dynamic operating conditions \cite{ayan2023comprehensive, shafik2024machine}. This motivates the hybrid design of QERS, where analytical scoring is augmented by a Random Forest–based fusion layer for robust PQC evaluation in real IoT and IIoT environments.

\subsection{Research Gap}
Although MQTT, HTTP, and HTTPS have been studied individually or in pairs for performance and PQC overhead, no previous work provides a unified, normalized metric to evaluate PQC resilience across protocols in constrained IoT/IIoT devices \cite{Silva16032021}. The Quantum Encryption Resilience Score (QERS) addresses this gap by integrating latency, packet loss, CPU usage, energy consumption, RSSI, key size, and cryptographic overhead into a single reproducible score for IoT/IIoT deployment.

\begin{itemize}
    \item \textit{Evaluation of Performance, Energy, and Computation Costs of Quantum-Attack Resilient Encryption Algorithms for Embedded Devices} \cite{PekPeterCh, halak2024securityassessmenttoolquantum}: Analyzes CPU, memory, and energy overhead of PQC on embedded devices.
    \item \textit{Post-Quantum Encryption Algorithms} \cite{PekPeterCh}: Reviews PQC algorithms and highlights efficiency and protocol-level trade-offs.
\end{itemize}

% --- Methodology ---
\section{Methodology (QERS Formulation)}

The Quantum Encryption Resilience Score (QERS) is formulated as a multi-criteria evaluation instrument for assessing the operational impact of Post-Quantum Cryptography (PQC) on constrained communication systems. The methodology follows a Multi-Criteria Decision Analysis (MCDA) structure combined with a Multi-Criteria Decision Making (MCDM) aggregation model, consistent with established evaluation frameworks used in protocol benchmarking and security assessment \cite{10.1007/978-981-32-9690-9_71, belton2002multiple, Ameyed2023}.
\vspace{0.1 cm}

Unlike isolated cryptographic benchmarks that focus solely on execution time or memory usage \cite{cryptography9020032}. QERS evaluates PQC in end-to-end communication contexts on traditional processors, explicitly incorporating protocol behavior observed in MQTT, HTTP, and HTTPS deployments on constrained devices \cite{grgic2016web}. This approach reflects prior benchmarking methodologies that emphasize system-level indicators such as latency, packet reliability,
energy consumption, and signal stability
\cite{iot6040062, metrics2020009, lall2025reviewcollectionmetricsbenchmarks}.
\vspace{0.1 cm}

\subsection{Metric Definition}
Each protocol–algorithm configuration is evaluated using the following metrics, selected to capture both performance penalties and security characteristics under PQC-enabled communication:

\begin{table}[H]
\centering
\caption{Mapping of MCDM Symbols to QERS Metrics}
\label{tab:mcdm_qers_mapping}
\scriptsize
\begin{tabular}{|c|p{0.70\columnwidth}|}
\hline
\textbf{Symbol} & \textbf{Description} \\ \hline

QERS & Quantum Encryption Resilience Score (normalized to 0--100) \\ \hline

$X_{\min}, X_{\max}$ & Minimum and maximum observed values used for normalization \\ \hline

$X_{\text{norm}}$ & Min--max normalized metric value \\ \hline

$x_{ij}$ & Value of criterion $j$ for alternative $i$ \\ \hline

$r_{ij}$ & Normalized form of $x_{ij}$ \\ \hline

$w_i$ & Weight assigned to criterion $i$ \\ \hline

$\sum_i w_i = 1$ & Weight constraint for performance criteria 
($L$, $J$, $P_{\text{loss}}$, $C$, $E$) \\ \hline

$\sum_j w_j = 1$ & Weight constraint for security criteria 
($K$, $R$, $P_r$, $Co$) \\ \hline

$A$ & Set of evaluated alternatives (MQTT, HTTP, HTTPS) \\ \hline

$X$ & Set of evaluation criteria \\ \hline

$\alpha$--$\eta$ & Decision weighting coefficients in QERS \\ \hline

$L$ & End-to-end latency \\ \hline

$J$ & Packet jitter \\ \hline

$P_{\text{loss}}$ & Packet loss ratio \\ \hline

$C$ & CPU utilization \\ \hline

$E$ & Energy consumption \\ \hline

$R$ & RSSI (signal strength indicator) \\ \hline

$K$ & Cryptographic key size (bytes) \\ \hline

$Co$ & Cryptographic overhead (memory, bandwidth, computation) \\ \hline

$P_r$ & Proven cryptographic resistance level \\ \hline

\end{tabular}

\vspace{4pt}
\begin{minipage}{0.95\linewidth}
\footnotesize \textit{Sources:} \cite{boo1, book, belton2002multiple}
\end{minipage}
\end{table}

These metrics align with protocol-level measurements reported in computer systems, IoT, and IIoT benchmarking literature, where MQTT emphasizes low latency and reliability, HTTP incurs higher payload overhead, and HTTPS introduces additional cryptographic and handshake costs
\cite{smartcities8040116, 10.1145/3716368.3735199, electronics14214234, jara2023comparative}.

\subsection{Metric Normalization}

To enable aggregation across heterogeneous measurement scales, all metrics are normalized using min--max scaling, a standard technique in MCDA-based composite indicators:
\vspace{0.1 cm}

\begin{equation}
X_{\text{norm}} =
MS \cdot \frac{X - X_{\min}}{X_{\max} - X_{\min}} \nonumber
\label{eq:minmax}
\end{equation}

\vspace{0.1 cm}

where $X$ denotes the observed metric value, $X_{\min}$ and $X_{\max}$ are derived from the experimental dataset, and $MS$ represents the maximum score (set to 100). Dataset-based bounds ensure that normalization reflects realistic operating conditions rather than theoretical extremes \cite{boo1, book, belton2002multiple}.
\vspace{0.1 cm}

\subsection{QERS Evaluation Layers}

QERS is computed using three hierarchical evaluation layers, enabling progressively deeper analysis depending on deployment requirements.
\vspace{0.1 cm}

\subsubsection{Basic QERS}
Basic QERS provides a rapid comparison of PQC impact across protocols by considering latency, cryptographic overhead, and packet loss:
\vspace{0.1 cm}

\begin{equation}
\text{QERS}_{\text{basic}} =
MS - \left(
\alpha L_{\text{norm}} +
\beta O_{\text{norm}} +
\gamma P_{\text{loss,norm}}
\right). \nonumber
\end{equation}
\vspace{0.1 cm}

This formulation is particularly suitable for contrasting lightweight protocols such as MQTT against HTTP and HTTPS, where packet loss and handshake overhead dominate performance behavior
\cite{gentile2024network,ghotbou2021comparing}.
\vspace{0.1 cm}

\subsubsection{Tuned QERS}
Tuned QERS extends the Basic formulation by incorporating system and environmental factors:
\vspace{0.1 cm}

\begin{align}
\text{QERS}_{\text{tuned}} = MS
&- \left(
\alpha L_{\text{norm}} +
\beta O_{\text{norm}} +
\gamma P_{\text{loss,norm}} \right. \nonumber \\
&\left.
+ \delta C_{\text{norm}} +
\zeta E_{\text{norm}} +
\eta K_{\text{norm}}
\right)
+ \epsilon R_{\text{norm}} . \nonumber
\end{align}
\vspace{0.1 cm}

Including RSSI reflects wireless stability effects observed in Wi-Fi based IoT deployments, where signal degradation directly impacts retransmissions and energy consumption
\cite{sadowski2018rssi, hernandez2020lightweight}.
\vspace{0.1 cm}

\subsubsection{Fusion QERS}
Fusion QERS separates performance penalties and security gains into two subscores.
\vspace{0.1 cm}

\paragraph{Performance Subscore}
\begin{equation}
P = \sum_i w_i C_i^{\text{norm}}, \quad \sum_i w_i = 1, \nonumber
\end{equation}

where $C_i \in \{L, J, P_{\text{loss}}, E, C\}$. 

\paragraph{Security Subscore}
\begin{equation}
S = \sum_j w_j B_j^{\text{norm}}, \quad \sum_j w_j = 1, \nonumber
\end{equation}

where $B_j \in \{K, R, P_r, Co\}$.

\paragraph{Fusion Score}
\begin{equation}
\text{QERS}_{\text{fusion}} =
\alpha (MS - P) + \beta S, \quad \alpha + \beta = 1. \nonumber
\end{equation}
\vspace{0.1 cm}

Fusion QERS is computed analytically using weighted performance and security subscores. No machine learning model is used in the current implementation; the Fusion score represents a deterministic multi-criteria
aggregation based on measured protocol behavior.This formulation ensures that reduced latency, CPU usage, and energy consumption increase the final score, while larger key sizes and higher proven resistance contribute positively to security resilience, in line with composite evaluation models used in IoT and IIoT security assessment \cite{boo1, 10763879, Ameyed2023}.
\vspace{0.1 cm}

\subsubsection{Weighting Strategy and Design Rationale}
The weighting coefficients $(\alpha, \beta, \gamma, \delta, \epsilon, \zeta, \eta)$ define the relative importance of performance and security metrics in the QERS
framework. Weights were selected using a Multi-Criteria Decision Analysis (MCDA) principle to reflect operational priorities while avoiding dominance by any
single metric \cite{boo1, belton2002multiple, book}.
\vspace{0.1 cm}

Latency, cryptographic overhead, packet loss, CPU usage, energy consumption, and key size are treated as cost criteria, while RSSI is treated as a benefit
criterion because higher signal strength improves link stability and reduces retransmissions \cite{boo1, book}. The same weight set was applied to MQTT, HTTP, and HTTPS to ensure protocol differences arise from measured behavior rather than parameter tuning.
\vspace{0.1 cm}

Fusion QERS separates performance penalties ($P$) and security benefits ($S$), allowing the framework to model the trade-off between efficiency and cryptographic robustness in post-quantum IoT systems \cite{wang2025review}.
\vspace{0.1 cm}

\begin{table}[h!]
\centering
\caption{Baseline Weight Configuration for QERS Metrics}
\label{tab:qers_weights}
\scriptsize
\begin{tabular}{|c|l|c|c|}
\hline
\textbf{Symbol} & \textbf{Metric} & \textbf{Type} & \textbf{Weight} \\ \hline
$L$ & Latency & Cost & 0.25 \\ \hline
$O$ & Cryptographic overhead & Cost & 0.15 \\ \hline
$P_{\text{loss}}$ & Packet loss & Cost & 0.15 \\ \hline
$C$ & CPU utilization & Cost & 0.15 \\ \hline
$E$ & Energy consumption & Cost & 0.10 \\ \hline
$K$ & Key size (PQC bytes) & Cost & 0.10 \\ \hline
$R$ & RSSI (signal strength) & Benefit & 0.10 \\ \hline
\multicolumn{3}{|c|}{\textbf{Total}} & \textbf{1.00} \\ \hline
\end{tabular}
\end{table}
\vspace{0.1 cm}

The baseline weight configuration reflects a balanced IoT deployment scenario in which latency and reliability are prioritized, while CPU, energy, and cryptographic overhead remain significant but secondary constraints. RSSI is treated as a benefit criterion because stronger
signal quality improves packet delivery and reduces retransmissions. This weighting scheme follows standard MCDA practice for composite performance indicators \cite{boo1, belton2002multiple, book}.
\vspace{0.1 cm}

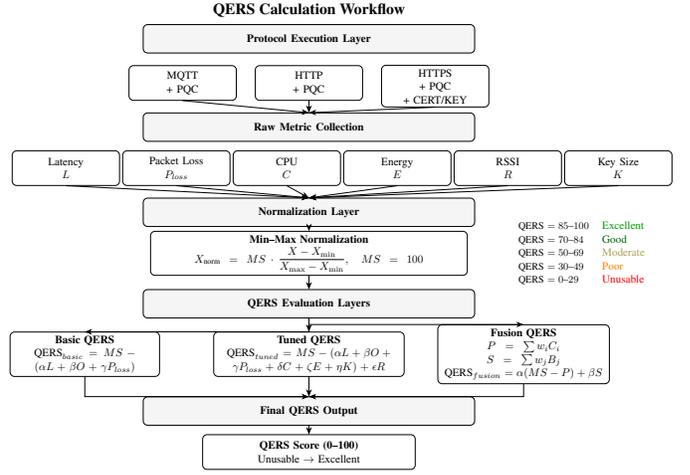
\begin{figure}[htbp]
\centering
\resizebox{\columnwidth}{!}{%
\begin{tikzpicture}[
    box/.style={rectangle, draw, rounded corners,
        minimum width=3.4cm, minimum height=1.1cm, align=center},
    layer/.style={rectangle, draw, rounded corners,
        minimum width=10.5cm, minimum height=0.9cm, align=center,
        fill=gray!8},
    arrow/.style={->, thick},
    >=Stealth
]

% ---------- TITLE ----------
\node at (6,5.1) {\Large \textbf{QERS Calculation Workflow}};

% ---------- PROTOCOL LAYER ----------
\node[layer] (proto) at (6,4.2) {\textbf{Protocol Execution Layer}};
\node[box] (mqtt) at (2,2.8) {MQTT \\ + PQC};
\node[box] (http) at (6,2.8) {HTTP \\ + PQC};
\node[box] (https) at (10,2.7) {HTTPS \\ + PQC \\ + CERT/KEY};

% ---------- METRICS LAYER ----------
\node[layer] (metrics) at (6,1.4) {\textbf{Raw Metric Collection}};
\node[box] (m1) at (-1.7,0.1) {Latency \\ $L$};
\node[box] (m2) at (1.8,0.1) {Packet Loss \\ $P_{loss}$};
\node[box] (m3) at (5.3,0.1) {CPU \\ $C$};
\node[box] (m4) at (8.8,0.1) {Energy \\ $E$};
\node[box] (m5) at (12.3,0.1) {RSSI \\ $R$};
\node[box] (m6) at (15.8,0.1) {Key Size \\ $K$};

% ---------- NORMALIZATION ----------
\node[layer] (normL) at (6,-1.3) {\textbf{Normalization Layer}};
\node[box, text width=9.8cm] (norm) at (6,-2.6)
{
\textbf{Min--Max Normalization}\\[4pt]
$
X_{\text{norm}} = MS \cdot
\dfrac{X - X_{\min}}{X_{\max} - X_{\min}},
\quad MS = 100
$
};

% ---------- QERS LAYERS ----------
\node[layer] (qersL) at (6,-4.2) {\textbf{QERS Evaluation Layers}};

\node[box, text width=4.5cm] (basic) at (-1.1,-5.8)
{
\textbf{Basic QERS}\\
$
\text{QERS}_{basic} =
MS - (\alpha L + \beta O + \gamma P_{loss})
$
};

\node[box, text width=5.8cm] (tuned) at (6,-5.8)
{
\textbf{Tuned QERS}\\
$
\text{QERS}_{tuned} =
MS - (\alpha L + \beta O + \gamma P_{loss}
+ \delta C + \zeta E + \eta K)
+ \epsilon R
$
};

\node[box, text width=5.2cm] (fusion) at (12.8,-5.8)
{
\textbf{Fusion QERS}\\
$P = \sum w_i C_i$ \\
$S = \sum w_j B_j$ \\
$
\text{QERS}_{fusion} =
\alpha(MS - P) + \beta S
$
};

% ---------- OUTPUT ----------
\node[layer] (outL) at (6,-7.6) {\textbf{Final QERS Output}};
\node[box, text width=6.5cm] (out) at (6,-8.9)
{
\textbf{QERS Score (0--100)}\\
Unusable $\rightarrow$ Excellent
};

% ---------- QERS READINESS LEGEND ----------
\node[align=left] at (14.7,-2.6) {
\begin{tabular}{@{}l l@{}}
\small QERS $=85$--$100$ & \textcolor{green!60!black}{Excellent} \\
\small QERS $=70$--$84$  & \textcolor{green!40!black}{Good} \\
\small QERS $=50$--$69$  & \textcolor{yellow!60!black}{Moderate} \\
\small QERS $=30$--$49$  & \textcolor{orange}{Poor} \\
\small QERS $=0$--$29$   & \textcolor{red}{Unusable} \\
\end{tabular}
};

% ---------- ARROWS ----------
\draw[arrow] (mqtt.south) -- (metrics.north);
\draw[arrow] (http.south) -- (metrics.north);
\draw[arrow] (https.south) -- (metrics.north);

\foreach \n in {m1,m2,m3,m4,m5,m6}
  \draw[arrow] (\n.south) -- (normL.north);

\draw[arrow] (normL.south) -- (norm.north);
\draw[arrow] (norm.south) -- (qersL.north);

\draw[arrow] (qersL.south) |- (basic.north);
\draw[arrow] (qersL.south) -- (tuned.north);
\draw[arrow] (qersL.south) |- (fusion.north);

\draw[arrow] (basic.south) |- (outL.north);
\draw[arrow] (tuned.south) -- (outL.north);
\draw[arrow] (fusion.south) |- (outL.north);

\draw[arrow] (outL.south) -- (out.north);

\end{tikzpicture}
}
\caption{End-to-end QERS calculation workflow showing protocol execution, metric collection, normalization, Basic/Tuned/Fusion evaluation, and final resilience score. HTTPS explicitly includes certificate handling and PQC-based key establishment.}
\label{fig:qers_workflow}
\end{figure}
\vspace{0.1 cm}

\subsection{TLS and Certificate Configuration}
For HTTPS measurements, the Raspberry Pi server was configured with a self-signed X.509 certificate and private key to enable TLS 1.3 connections. The ESP32 client validated the server certificate during the TLS handshake before initiating encrypted communication. This ensured that HTTPS measurements reflect realistic authentication, key exchange, and cryptographic overhead when PQC algorithms are used for secure communication.
\vspace{0.1 cm}

\begin{figure}[htbp]
\centering
\resizebox{\columnwidth}{!}{%
\begin{tikzpicture}[
    box/.style={rectangle, draw, rounded corners, minimum width=3.2cm, minimum height=1cm, align=center},
    layer/.style={rectangle, draw, dashed, rounded corners, minimum width=7.2cm, minimum height=1cm, align=center},
    arrow/.style={->, thick},
    >=Stealth
]

% ================== DEVICES ==================
\node[box] (esp) at (0,0) {ESP32-C6 \\ PQC Client};
\node[box] (pi) at (12,0) {Raspberry Pi CM4 \\ QERS Server};

% ================== PROTOCOL STACK ==================
\node[box] (mqtt) at (0,-2.2) {MQTT};
\node[box] (http) at (0,-4.2) {HTTP};
\node[box] (https) at (0,-6.2) {HTTPS + TLS};

% ================== PQC LAYER ==================
\node[box] (kem) at (5,-5.2) {Kyber \\ (PQC Key Exchange)};
\node[box] (sig) at (5,-7.0) {Dilithium \\ (PQC Signature)};

% ================== SERVER SIDE ==================
\node[box] (cert) at (9,-6.8) {Dilithium-Signed \\ X.509 Certificate};
\node[box] (metrics) at (12,-3.3) {Latency, CPU, RSSI \\ Key Size, TLS Time};
\node[box] (qers) at (12,-5.4) {QERS Engine \\ Basic / Tuned / Fusion};

% ================== RANGE CONTEXT ==================
\node[layer] (close) at (6,1.5) {Close Range (Near Wi-Fi AP)};
\node[layer] (far) at (6,-8.8) {Far Range (10 feet from AP)};

% ================== WIRELESS LINK ==================
\draw[arrow] (esp.east) -- (pi.west) node[midway, above] {Wi-Fi};

% ================== PROTOCOL FLOW ==================
\draw[arrow] (esp.south) -- (mqtt.north);
\draw[arrow] (mqtt.south) -- (http.north);
\draw[arrow] (http.south) -- (https.north);

% ================== PQC APPLIED TO ALL PROTOCOLS ==================
\draw[arrow] (mqtt.east) -- ++(2.5,0) |- (kem.north);
\draw[arrow] (http.east) -- ++(2.5,0) |- (kem.north);
\draw[arrow] (https.east) -- (kem.west);

\draw[arrow] (kem.south) -- (sig.north);
\draw[arrow] (sig.east) -- (cert.west);

% ================== SERVER PROCESSING ==================
\draw[arrow] (pi.south) -- (metrics.north);
\draw[arrow] (metrics.south) -- (qers.north);
\draw[arrow] (cert.north) -- (qers.south);

\end{tikzpicture}
}
\caption{End-to-end QERS workflow showing MQTT, HTTP, and HTTPS operating under Kyber-based PQC key exchange and Dilithium-signed authentication. All protocols are evaluated on an ESP32-C6 client communicating with a Raspberry Pi QERS server under close-range and far-range wireless conditions.}
\label{fig:qers_workflow}
\end{figure}
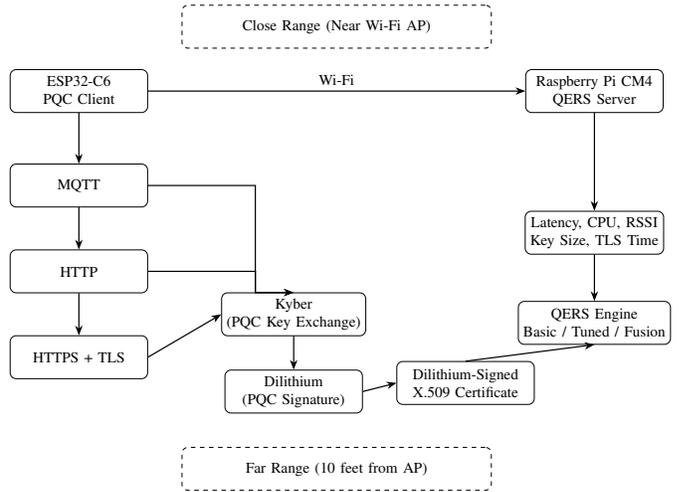

\subsection{Measurement Campaign}

Each protocol–PQC configuration (MQTT, HTTP, and HTTPS using Kyber and Dilithium) was measured continuously for a 12-hour period in both close-range and far-range wireless conditions. This long-duration acquisition produces a statistically stable representation of protocol behavior under realistic PQC load, allowing QERS to capture both transient and sustained performance effects. Metrics were sampled at fixed time intervals and recorded to CSV logs on the Raspberry\~Pi server, including latency, CPU utilization, RSSI, energy consumption, and QERS (Basic, Tuned, and Fusion) values. This long-duration acquisition allows the analysis to capture temporal variations in wireless conditions, cryptographic workload, and system stability that cannot be observed in short benchmark runs.

% -------------------------------------------------
\section{Results}
A continuous measurement campaign was conducted using an ESP32-C6 client  and a Raspberry Pi CM4 server, evaluating MQTT, HTTP, and HTTPS communication under post-quantum cryptographic (PQC) load. Measurements were recorded for latency, CPU utilization, RSSI, key size, and the derived QERS scores (Basic, Tuned, and Fusion). All values were normalized to a 0--100 scale.
\vspace{0.1 cm}

Two experimental scenarios were evaluated:
\vspace{0.1 cm}

\begin{itemize}
    \item \textbf{Scenario 1 (Close range):} ESP32 positioned close to the access point.
    \item \textbf{Scenario 2 (Far range):} ESP32 positioned approximately 10 feet from the access point.
\end{itemize}
\vspace{0.1 cm}

\subsection{Multi-Metric Heatmap Analysis}
\begin{figure}[htbp]
    \centering
    \includegraphics[width=0.45\textwidth]{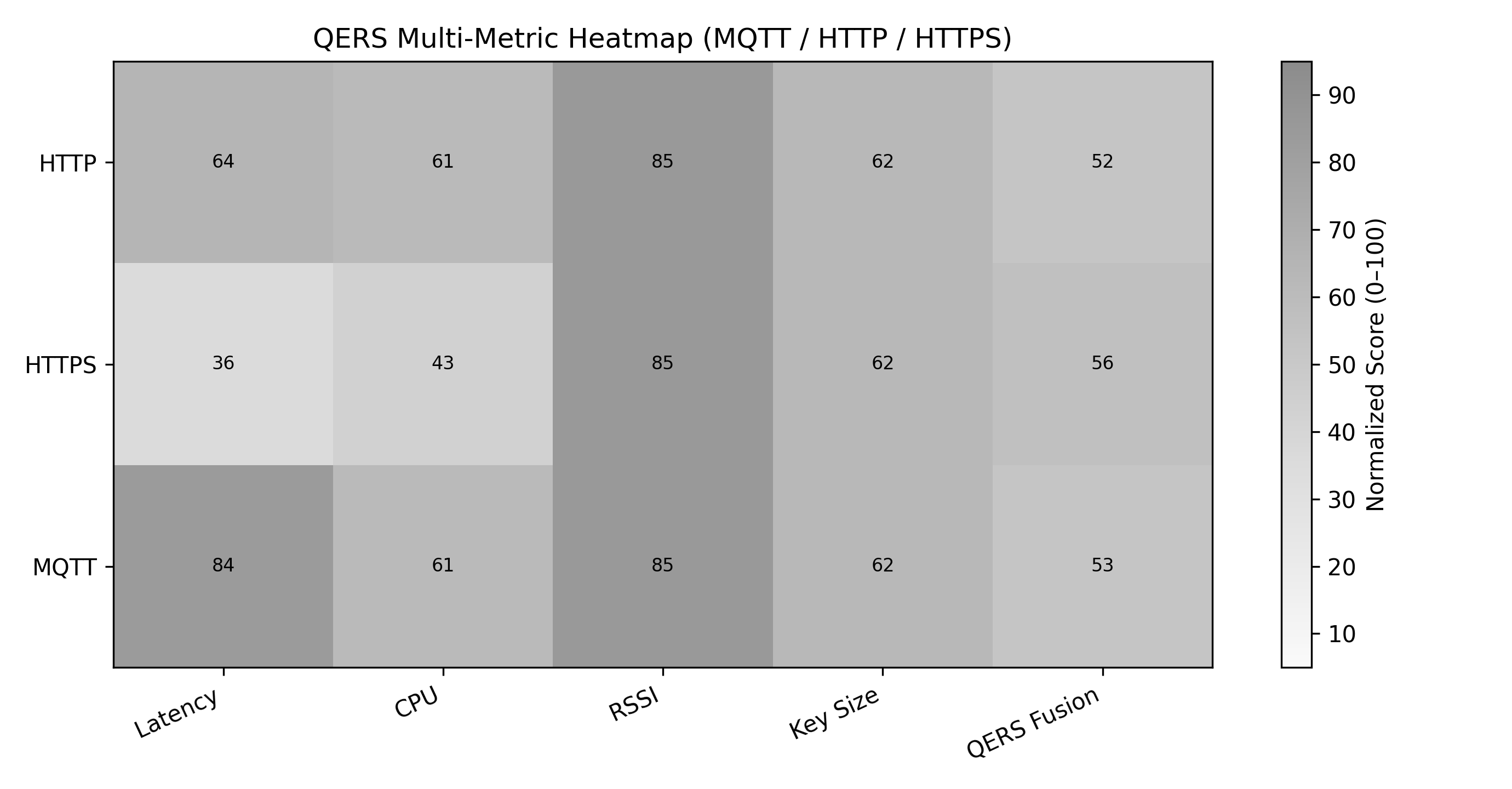}
    \caption{Scenario~1: Normalized QERS metrics (Latency, CPU, RSSI, Key Size, Fusion) for MQTT, HTTP, and HTTPS under Kyber–Dilithium PQC at close range.}
    \label{fig:heat1}
\end{figure}

\begin{figure}[htbp]
    \centering
    \includegraphics[width=0.45\textwidth]{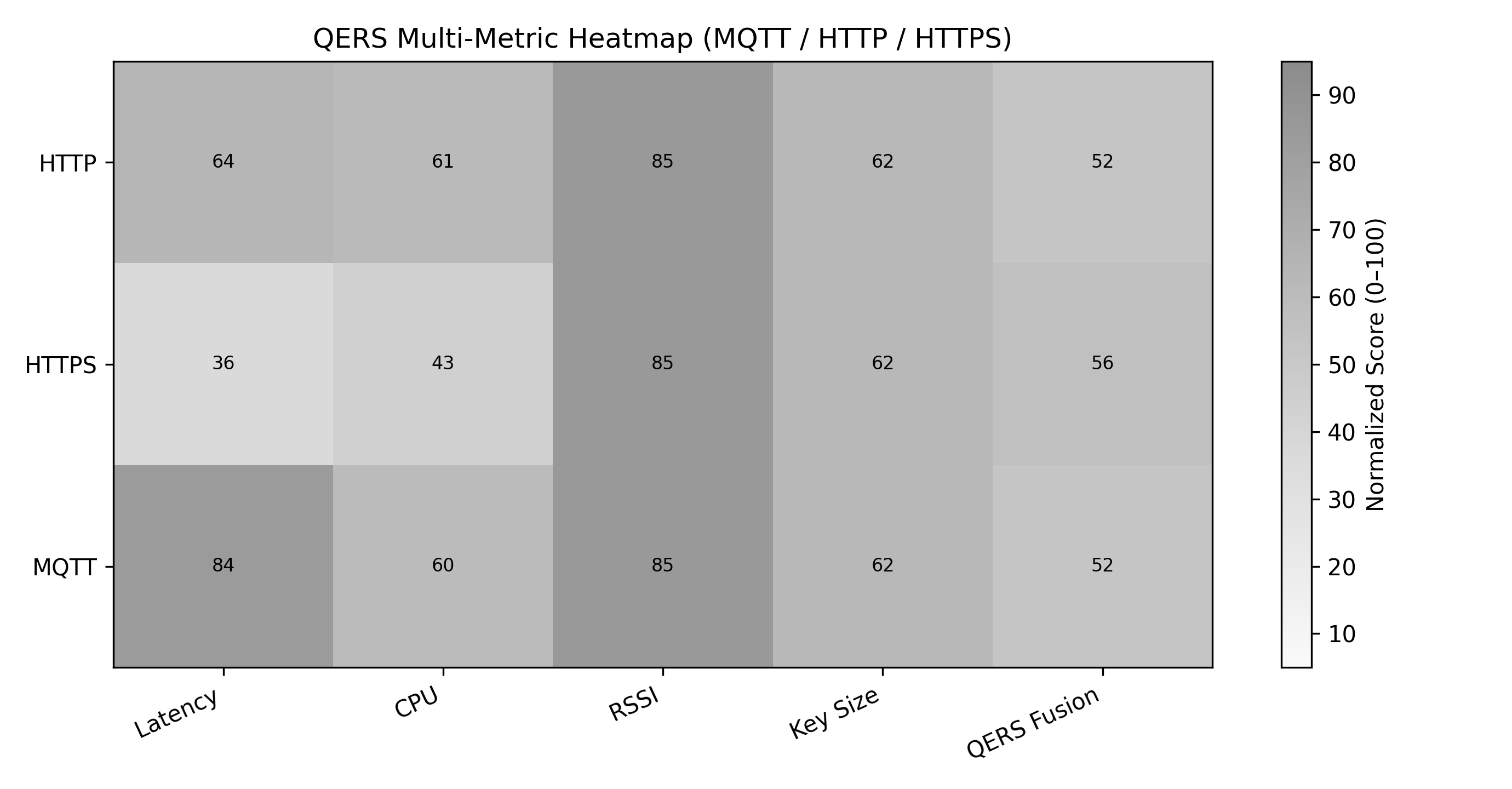}
    \caption{Scenario~2: QERS multi-metric heatmap (far range, normalized 0--100).}
    \label{fig:heat2}
\end{figure}

Figures~\ref{fig:heat1} and~\ref{fig:heat2} illustrate normalized multi-metric heatmaps for MQTT, HTTP, and HTTPS under close and far communication scenarios.
\vspace{0.1 cm}

\vspace{0.1 cm}

Although HTTPS exhibits the lowest Basic and Tuned QERS values due to the computational and handshake overhead introduced by TLS and post-quantum cryptography, it achieves the highest Fusion QERS in both measurement scenarios. This is because Fusion QERS explicitly rewards cryptographic strength, authentication, and transport-layer security, whereas Basic and Tuned QERS primarily reflect efficiency and resource consumption.
\vspace{0.1 cm}

Across both scenarios, MQTT consistently achieved the highest normalized latency scores, indicating lower communication delay under PQC load. HTTPS exhibits the lowest efficiency (Basic/Tuned QERS) but the highest security-weighted Fusion QERS, reflecting its stronger cryptographic and authentication guarantees. RSSI values remained comparable across protocols, confirming that observed differences primarily originate from protocol and cryptographic behavior rather than signal instability.
\vspace{0.1 cm}

\subsection{Representative Scatter Analysis}
\begin{figure}[htbp]
    \centering
    \includegraphics[width=0.45\textwidth]{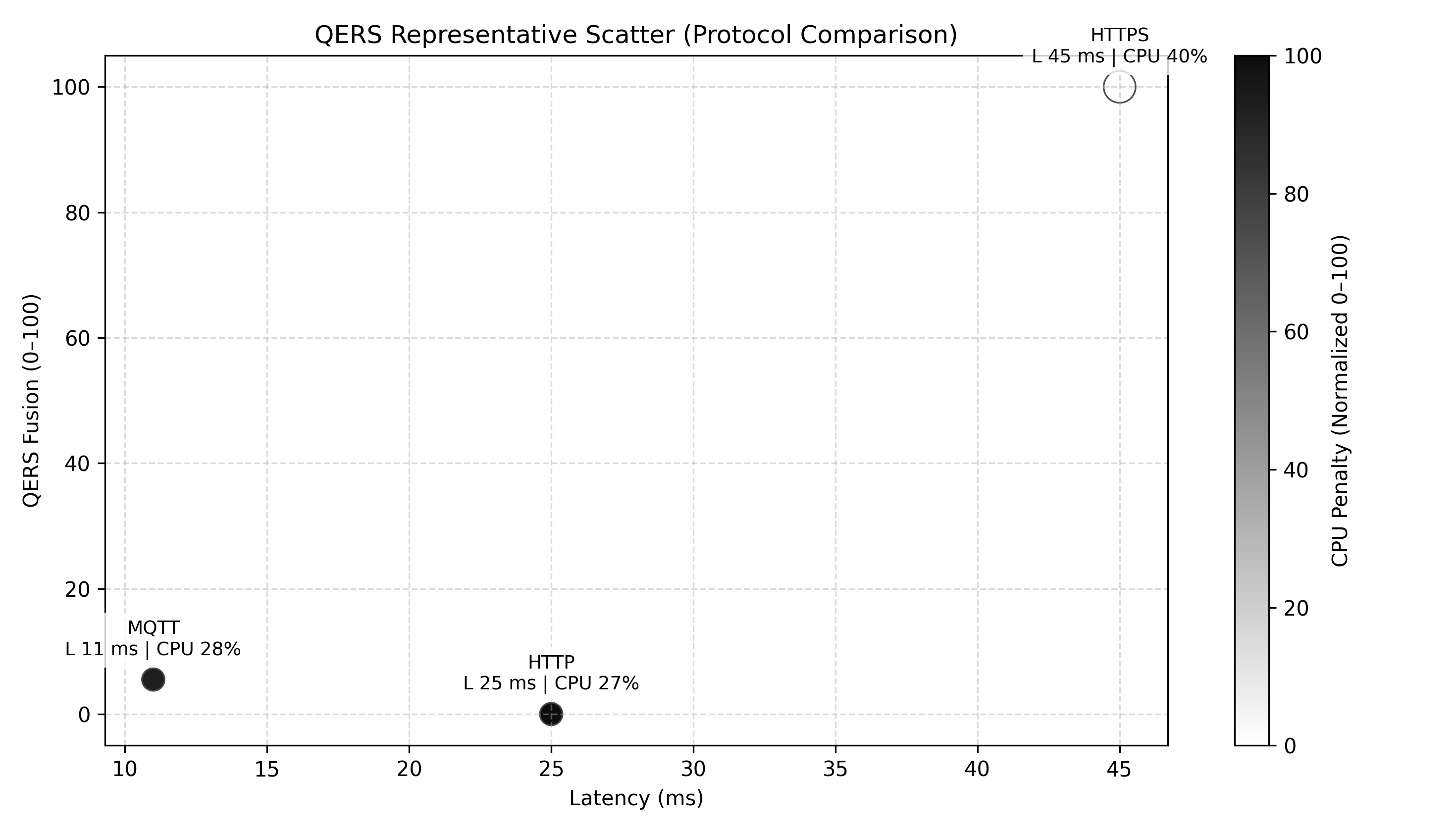}
    \caption{Scenario~1: Representative scatter plot (latency vs QERS Fusion).}
    \label{fig:scatter1}
\end{figure}

\begin{figure}[htbp]
    \centering
    \includegraphics[width=0.45\textwidth]{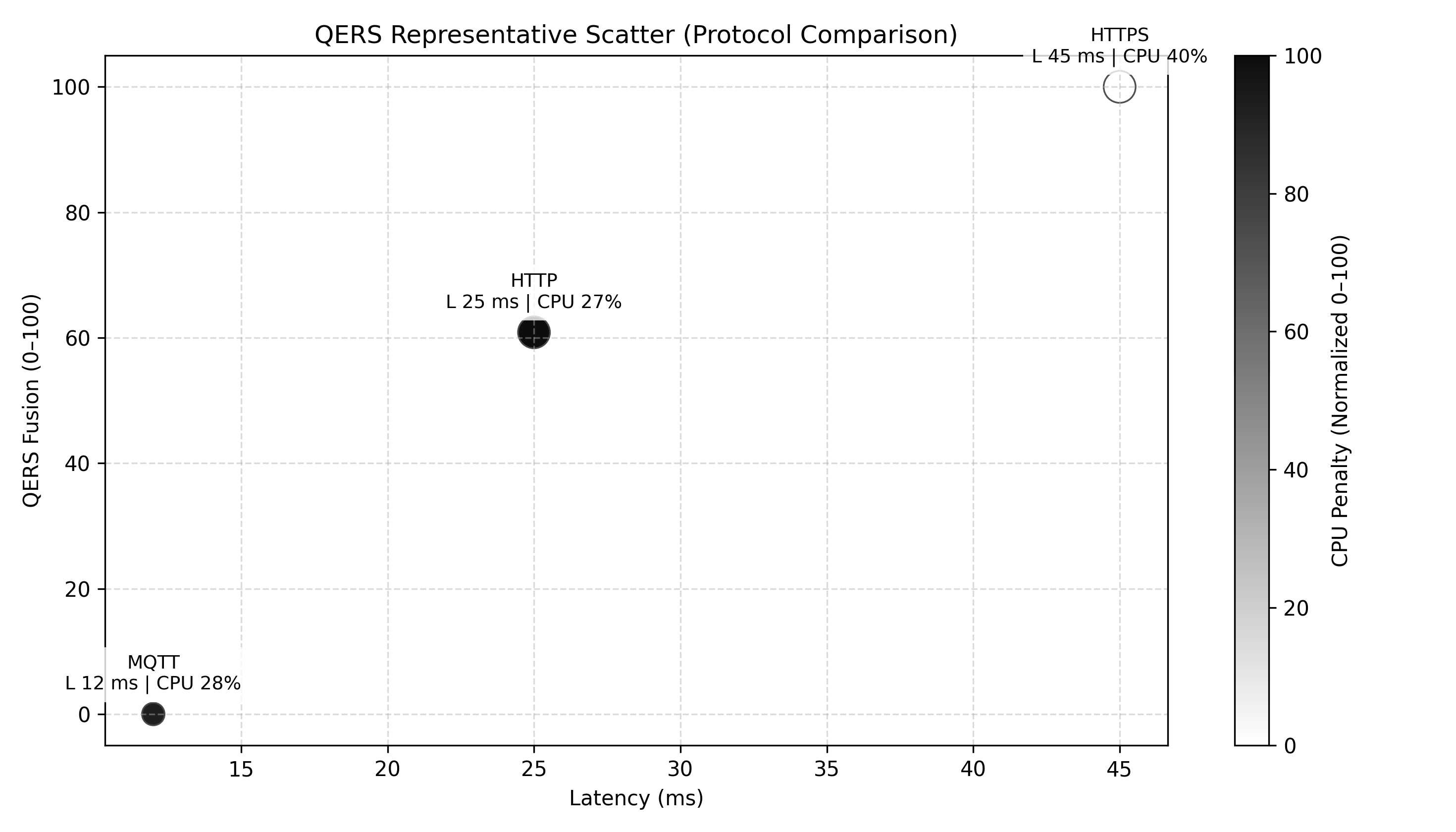}
    \caption{Scenario~2: Representative scatter plot (latency vs QERS Fusion).}
    \label{fig:scatter2}
\end{figure}

Figures~\ref{fig:scatter1} and~\ref{fig:scatter2} present representative scatter plots using median latency and Fusion QERS values per protocol. Bubble shading represents normalized CPU penalty.
\vspace{0.1 cm}

\vspace{0.1 cm}

MQTT occupies the low-latency, high-efficiency region in both scenarios. HTTP demonstrates intermediate performance, while HTTPS consistently shows increased latency and CPU utilization, leading to reduced Fusion QERS scores.
\vspace{0.1 cm}

\subsection{QERS Score Distribution}
\begin{figure}[htbp]
    \centering
    \includegraphics[width=0.45\textwidth]{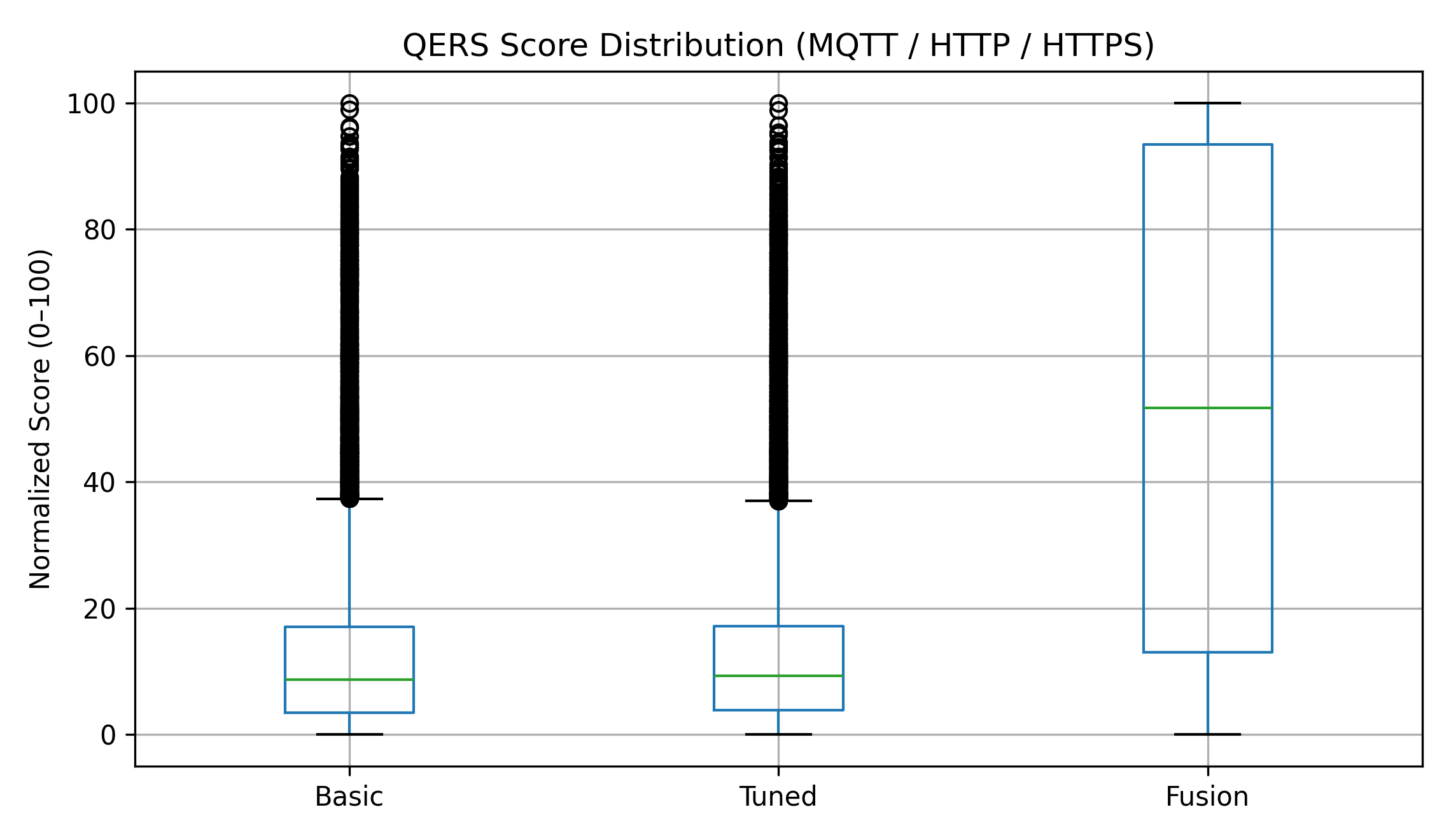}
    \caption{Scenario~1: QERS score distributions (close range).}
    \label{fig:box1}
\end{figure}

\begin{figure}[htbp]
    \centering
    \includegraphics[width=0.45\textwidth]{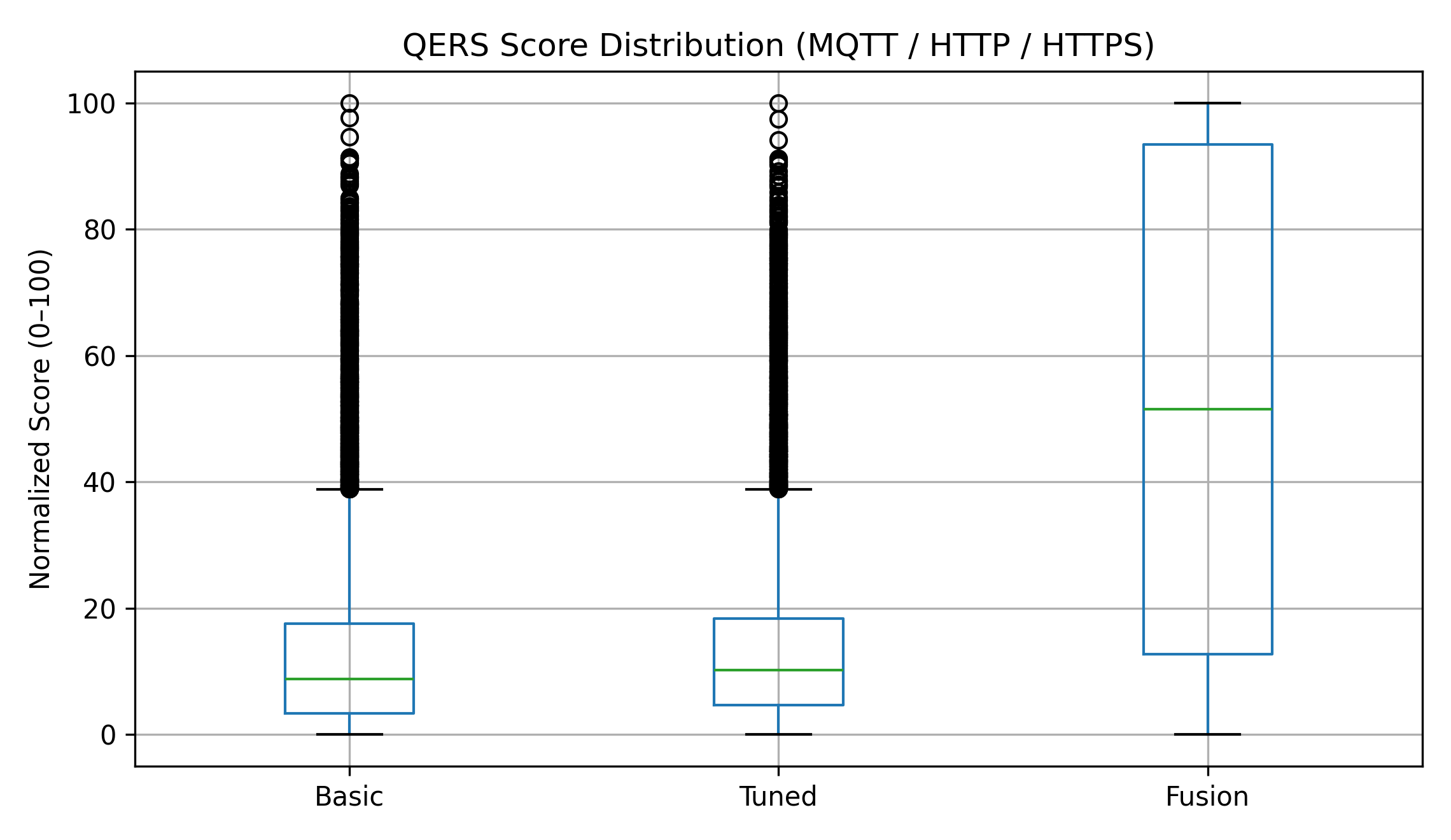}
    \caption{Scenario~2: QERS score distributions (far range).}
    \label{fig:box2}
\end{figure}

Figures~\ref{fig:box1} and~\ref{fig:box2} illustrate the distribution of Basic, Tuned, and Fusion QERS scores across the full measurement period.
\vspace{0.1 cm}

\vspace{0.1 cm}

Fusion QERS exhibits higher median values and reduced variability compared to Basic and Tuned scores, demonstrating its ability to smooth short-term fluctuations while preserving protocol-specific behavior.
\vspace{0.1 cm}

\begin{table}[H]
\centering
\caption{Average QERS scores across protocols and distance scenarios (normalized 0--100).}
\label{tab:qers_protocol_summary}
\begin{tabular}{lcccc}
\hline
\textbf{Scenario} & \textbf{Protocol} & \textbf{QERS$_{Basic}$} & \textbf{QERS$_{Tuned}$} & \textbf{QERS$_{Fusion}$} \\
\hline
Close & MQTT  & 61.39 & 60.76 & 50.87 \\
Close & HTTP  & 25.44 & 24.13 & 49.97 \\
Close & HTTPS & 11.06 & 11.67 & 54.65 \\
\hline
Far (10 ft) & MQTT  & 60.44 & 59.77 & 50.87 \\
Far (10 ft) & HTTP  & 24.22 & 22.92 & 50.09 \\
Far (10 ft) & HTTPS & 10.75 & 11.15 & 54.69 \\
\hline
\end{tabular}
\end{table}
\vspace{0.1 cm}

According to the QERS interpretation scale, Fusion scores around 50 correspond to a \textit{Moderate} readiness level. In this context, MQTT and HTTP provide the most efficient communication under PQC (high Basic and Tuned QERS), while HTTPS provides the strongest security posture (highest Fusion QERS), at the cost of higher latency and resource consumption. This indicates that while PQC-enabled communication is feasible, none of the tested protocol stacks yet achieves excellent readiness under the tested hardware constraints.

% --- Discussion ---
\section{Discussion}
As summarized in Table~\ref{tab:qers_protocol_summary}, MQTT achieves the highest Basic and Tuned QERS, while HTTPS achieves the highest Fusion QERS.
\vspace{0.1 cm}

The experimental results confirm that protocol choice significantly influences the viability of deploying post-quantum cryptography on resource-constrained IoT and IIoT platforms \cite{asif2021post, abdelhafeez2019study}. While all three protocols operate correctly under PQC load, their performance and security trade-offs differ substantially. MQTT consistently achieved the highest Basic and Tuned QERS values and high Fusion scores \cite{hong2023performance, rampazzo2023assessment}. Its persistent connections, lightweight headers, and publish--subscribe architecture reduce both latency and cryptographic overhead, making it well suited for PQC-enabled constrained environments
\cite{detti2020sub}. MQTT scores highest in Basic and Tuned QERS and achieves competitive Fusion QERS, although it provides less built-in cryptographic assurance than HTTPS.
\vspace{0.1 cm}

HTTP demonstrated moderate QERS performance. Although simpler than HTTPS, its stateless request--response model introduces repeated protocol and cryptographic overhead, which becomes increasingly visible under larger PQC key sizes \cite{akpakwu2025congestion, leggieri2013interoperability}. HTTPS exhibited the lowest Basic and Tuned QERS scores, reflecting the high computational and communication cost introduced by TLS handshakes, certificate exchange, and PQC key encapsulation \cite{scott2023tls, nofal2019comprehensive}.
\vspace{0.1 cm}

However, HTTPS achieved the highest Fusion QERS values, indicating that while it is the least efficient protocol, it provides the strongest overall security-weighted resilience under PQC. These effects are further amplified under increased distance, which highlights the sensitivity of PQC-enabled HTTPS to environmental conditions \cite{yang2013rssi}. Fusion QERS is intentionally security-biased; therefore HTTPS dominates when security weight $\beta$ is large. The distance comparison further validates the design of QERS. Moving the ESP32 farther from the access point produced a consistent reduction in QERS Fusion scores across all protocols \cite{el2018performance, lui2011differences}. This confirms that QERS captures environmental stress in addition to cryptographic and protocol-induced overhead \cite{pereira2017performance, 7498318}. Importantly, Fusion QERS does not eliminate protocol differences. Instead, it provides a stable composite indicator that preserves real system behavior while reducing sensitivity to transient measurement noise \cite{sobral2016composite, krishnamurthi2020overview}.
\vspace{0.1 cm}

% --- Conclusion ---
\section{Conclusion}

This research presents an experimental evaluation of the Quantum Encryption Resilience Score (QERS) applied to MQTT, HTTP and HTTPS communication protocols under post-quantum cryptographic workloads. Using an ESP32-C6 client and an ARM-based Raspberry Pi CM4 server, QERS was evaluated on representative constrained computing platforms. QERS was shown to effectively quantify protocol-level trade-offs by integrating latency, CPU utilization, RSSI, Key-Size, and cryptographic overhead into a single normalized metric. 
\vspace{0.1 cm}

The results demonstrate that MQTT offers the highest resilience under PQC constraints, followed by HTTP, while HTTPS incurs the largest performance penalties.
Distance-induced degradation further confirms that QERS captures both computational and environmental influences on secure communication.
\vspace{0.1 cm}

In Conclusion, QERS provides a reproducible, protocol-aware evaluation instrument that enables systematic comparison of PQC readiness in constrained computer systems IoT and IIoT environments. Future work will extend QERS with additional metrics such as packet loss variance, temperature, and long-term energy profiling, and will explore cross-platform validation on current computer hardware.

% --- References ---

\bibliographystyle{IEEEtran}
\bibliography{bibfil}

@misc{sen2025securityprivacymanagementiot,
      title={Security and Privacy Management of IoT Using Quantum Computing}, 
      author={Jaydip Sen},
      year={2025},
      eprint={2511.03538},
      archivePrefix={arXiv},
      primaryClass={cs.CR},
      url={https://arxiv.org/abs/2511.03538}, 
}

@article{kappler2022post,
  title={Post-quantum cryptography: An introductory overview and implementation challenges of quantum-resistant algorithms},
  author={K{\"a}ppler, Sherdel A and Schneider, Bettina},
  journal={Proceedings of the Society},
  volume={84},
  pages={61--71},
  year={2022}
}

@article{Joseph2022,
  author    = {David Joseph and Rafael Misoczki and Marc Manzano and Joe Tricot and Fernando Dominguez Pinuaga and Olivier Lacombe and Stefan Leichenauer and Jack Hidary and Phil Venables and Royal Hansen},
  title     = {Transitioning organizations to post-quantum cryptography},
  journal   = {Nature},
  year      = {2022},
  volume    = {605},
  number    = {7909},
  pages     = {237--243},
  doi       = {10.1038/s41586-022-04623-2},
  url       = {https://doi.org/10.1038/s41586-022-04623-2},
  issn      = {1476-4687}
}

@article{bankar2025next,
  title={Next-Generation Embedded Systems: Innovations in IoT Integration, Real-Time Processing, and Energy Efficiency},
  author={Bankar, Shripad and Balkrishna, Gururaj and Lokare, Amit},
  journal={Authorea Preprints},
  year={2025},
  publisher={Authorea}
}

@INPROCEEDINGS{11059522,
  author={Alnaseri, Omar and Himeur, Yassine and Atalla, Shadi and Mansoor, Wathiq},
  booktitle={2025 International Wireless Communications and Mobile Computing (IWCMC)}, 
  title={Complexity of Post-Quantum Cryptography in Embedded Systems and Its Optimization Strategies}, 
  year= {2025},
  volume={},
  number={},
  pages={776-781},
  doi={10.1109/IWCMC65282.2025.11059522}}

@ARTICLE{10988807,
  author={Turnip, Togu Novriansyah and Andersen, Birger and Vargas-Rosales, Cesar},
  journal={IEEE Communications Surveys \& Tutorials}, 
  title={Towards 6G Authentication and Key Agreement Protocol: A Survey on Hybrid Post Quantum Cryptography}, 
  year={2025},
  volume={},
  number={},
  pages={1-1},
  doi={10.1109/COMST.2025.3567439}}

@article{rahman2025comprehensive,
  title={A Comprehensive Review of M2M Communication Protocols},
  author={Rahman, Md Hafizur and Naderuzzaman, M},
  journal={Open Access Journal on Engineering Applications},
  volume={1},
  number={01},
  pages={1--13},
  year={2025},
  publisher={Open Access Journal on Engineering Applications}
}

@ARTICLE{10382535,
  author={Halak, Basel and Gibson, Thomas and Henley, Millicent and Botea, Cristin-Bianca and Heath, Benjamin and Khan, Sayedur},
  journal={IEEE Access}, 
  title={Evaluation of Performance, Energy, and Computation Costs of Quantum-Attack Resilient Encryption Algorithms for Embedded Devices}, 
  year={2024},
  volume={12},
  number={},
  pages={8791-8805},
  doi={10.1109/ACCESS.2024.3350775}}

@misc{lall2025reviewcollectionmetricsbenchmarks,
      title={A Review and Collection of Metrics and Benchmarks for Quantum Computers: definitions, methodologies and software}, 
      author={Deep Lall and Abhishek Agarwal and Weixi Zhang and Lachlan Lindoy and Tobias Lindström and Stephanie Webster and Simon Hall and Nicholas Chancellor and Petros Wallden and Raul Garcia-Patron and Elham Kashefi and Viv Kendon and Jonathan Pritchard and Alessandro Rossi and Animesh Datta and Theodoros Kapourniotis and Konstantinos Georgopoulos and Ivan Rungger},
      year={2025},
      eprint={2502.06717},
      archivePrefix={arXiv},
      primaryClass={quant-ph},
      url={https://arxiv.org/abs/2502.06717}, 
}

@inproceedings{ulanov2024multi,
  title={Multi-criteria evaluation of reference terms requirements for development work performance},
  author={Ulanov, DV},
  booktitle={E3S Web of Conferences},
  volume={515},
  pages={01007},
  year={2024},
  organization={EDP Sciences}
}

@INPROCEEDINGS{11132566,
  author={Kundu, Ashish and Kompella, Ramana},
  booktitle={2025 62nd ACM/IEEE Design Automation Conference (DAC)}, 
  title={Quantum-Resistant Security: PQC Readiness and Research Challenges (Invited)}, 
  year={2025},
  volume={},
  number={},
  pages={1-4},
  doi={10.1109/DAC63849.2025.11132566}}

@misc{lopez2025evaluatingpostquantumcryptographicalgorithms,
      title={Evaluating Post-Quantum Cryptographic Algorithms on Resource-Constrained Devices}, 
      author={Jesus Lopez and Viviana Cadena and Mohammad Saidur Rahman},
      year={2025},
      eprint={2507.08312},
      archivePrefix={arXiv},
      primaryClass={cs.CR},
      url={https://arxiv.org/abs/2507.08312}, 
}

@article{KOTANGALE2025103643,
title = {An Iterative quantum chaotic security method design for big data deployments using a multi-model framework for key generation, threat modelling, and anomaly-resilient validation},
journal = {MethodsX},
volume = {15},
pages = {103643},
year = {2025},
issn = {2215-0161},
doi = {https://doi.org/10.1016/j.mex.2025.103643},
url = {https://www.sciencedirect.com/science/article/pii/S221501612500487X},
author = {Archana Kotangale and Meesala Sudhir Kumar},
}

@Article{app11114879,
AUTHOR = {Silva, Daniel and Carvalho, Liliana I. and Soares, José and Sofia, Rute C.},
TITLE = {A Performance Analysis of Internet of Things Networking Protocols: Evaluating MQTT, CoAP, OPC UA},
JOURNAL = {Applied Sciences},
VOLUME = {11},
YEAR = {2021},
NUMBER = {11},
ARTICLE-NUMBER = {4879},
URL = {https://www.mdpi.com/2076-3417/11/11/4879},
ISSN = {2076-3417},
DOI = {10.3390/app11114879}
}

@Article{s22228852,
AUTHOR = {Palmese, Fabio and Redondi, Alessandro E. C. and Cesana, Matteo},
TITLE = {Adaptive Quality of Service Control for MQTT-SN},
JOURNAL = {Sensors},
VOLUME = {22},
YEAR = {2022},
NUMBER = {22},
ARTICLE-NUMBER = {8852},
URL = {https://www.mdpi.com/1424-8220/22/22/8852},
PubMedID = {36433448},
ISSN = {1424-8220},
DOI = {10.3390/s22228852}
}

@INPROCEEDINGS{8914552,
  author={Moraes, Thays and Nogueira, Bruno and Lira, Victor and Tavares, Eduardo},
  booktitle={2019 IEEE International Conference on Systems, Man and Cybernetics (SMC)}, 
  title={Performance Comparison of IoT Communication Protocols}, 
  year={2019},
  volume={},
  number={},
  pages={3249-3254},
  doi={10.1109/SMC.2019.8914552}}

@INPROCEEDINGS{8971097,
  author={Sadeq, Abdulrahman Sameer and Hassan, Rosilah and Al-rawi, Salah Sleibi and Jubair, Ahmed Mahdi and Aman, Azana Hafizah Mohd},
  booktitle={2019 International Conference on Cybersecurity (ICoCSec)}, 
  title={A Qos Approach For Internet Of Things (Iot) Environment Using Mqtt Protocol}, 
  year={2019},
  volume={},
  number={},
  pages={59-63},
  doi={10.1109/ICoCSec47621.2019.8971097}}

@INPROCEEDINGS{8765692,
  author={Toldinas, Jevgenijus and Lozinskis, Borisas and Baranauskas, Edgaras and Dobrovolskis, Algirdas},
  booktitle={2019 23rd International Conference Electronics}, 
  title={MQTT Quality of Service versus Energy Consumption}, 
  year={2019},
  volume={},
  number={},
  pages={1-4},
  doi={10.1109/ELECTRONICS.2019.8765692}}

@Article{jcp3030021,
AUTHOR = {Samandari, Juliet and Gritti, Clémentine},
TITLE = {Post-Quantum Authentication in the MQTT Protocol},
JOURNAL = {Journal of Cybersecurity and Privacy},
VOLUME = {3},
YEAR = {2023},
NUMBER = {3},
PAGES = {416--434},
URL = {https://www.mdpi.com/2624-800X/3/3/21},
ISSN = {2624-800X},
DOI = {10.3390/jcp3030021}
}

@article{Kumar2025,
  author    = {N. Rajesh Kumar and R. Bala Krishnan and Subramaniyaswamy Vairavasundaram and G. Manikandan and Indragandhi V and Logesh Ravi},
  title     = {An IoT based remote medical diagnosis system using one time pad cipher over MQTT protocol},
  journal   = {Scientific Reports},
  year      = {2025},
  volume    = {15},
  number    = {1},
  pages     = {42117},
  doi       = {10.1038/s41598-025-26208-5},
  url       = {https://doi.org/10.1038/s41598-025-26208-5},
  issn      = {2045-2322}
}

@INPROCEEDINGS{7184865,
  author={Pérez, Juan Luis and Carrera, David},
  booktitle={2015 IEEE First International Conference on Big Data Computing Service and Applications}, 
  title={Performance Characterization of the Servioticy API: An IoT-as-a-Service Data Management Platform}, 
  year={2015},
  volume={},
  number={},
  pages={62-71},
  doi={10.1109/BigDataService.2015.58}}

@INPROCEEDINGS{8620130,
  author={Ismail, Ahmed A. and Hamza, Haitham S. and Kotb, Amira M.},
  booktitle={2018 IEEE Global Conference on Internet of Things (GCIoT)}, 
  title={Performance Evaluation of Open Source IoT Platforms}, 
  year={2018},
  volume={},
  number={},
  pages={1-5},
  doi={10.1109/GCIoT.2018.8620130}}

@INPROCEEDINGS{8400067,
  author={Bziuk, Wolfgang and Phung, Cao Vien and Dizdarević, Jasenka and Jukan, Admela},
  booktitle={2018 41st International Convention on Information and Communication Technology, Electronics and Microelectronics (MIPRO)}, 
  title={On HTTP performance in IoT applications: An analysis of latency and throughput}, 
  year={2018},
  volume={},
  number={},
  pages={0350-0355},
  doi={10.23919/MIPRO.2018.8400067}}

@inproceedings{10.1145/3465481.3465747,
author = {Paul, Sebastian and Schick, Felix and Seedorf, Jan},
title = {TPM-Based Post-Quantum Cryptography: A Case Study on Quantum-Resistant and Mutually Authenticated TLS for IoT Environments},
year = {2021},
isbn = {9781450390514},
publisher = {Association for Computing Machinery},
address = {New York, NY, USA},
url = {https://doi.org/10.1145/3465481.3465747},
doi = {10.1145/3465481.3465747},
booktitle = {Proceedings of the 16th International Conference on Availability, Reliability and Security},
articleno = {3},
numpages = {10},
location = {Vienna, Austria},
series = {ARES '21}
}

@article{CAIAZZA2024101871,
title = {Energy consumption of smartphones and IoT devices when using different versions of the HTTP protocol},
journal = {Pervasive and Mobile Computing},
volume = {97},
pages = {101871},
year = {2024},
issn = {1574-1192},
doi = {https://doi.org/10.1016/j.pmcj.2023.101871},
url = {https://www.sciencedirect.com/science/article/pii/S1574119223001293},
author = {Chiara Caiazza and Valerio Luconi and Alessio Vecchio},
}

@Article{s25196042,
AUTHOR = {Krawiec, Jerzy and Wybraniak-Kujawa, Martyna and Jacyna-Gołda, Ilona and Kotylak, Piotr and Panek, Aleksandra and Wojtachnik, Robert and Siedlecka-Wójcikowska, Teresa},
TITLE = {Energy Footprint and Reliability of IoT Communication Protocols for Remote Sensor Networks},
JOURNAL = {Sensors},
VOLUME = {25},
YEAR = {2025},
NUMBER = {19},
ARTICLE-NUMBER = {6042},
URL = {https://www.mdpi.com/1424-8220/25/19/6042},
PubMedID = {41094866},
ISSN = {1424-8220},
DOI = {10.3390/s25196042}
}

@article{Silva16032021,
author = {Edgar M. Silva and Ricardo Jardim-Goncalves},
title = {Cyber-Physical Systems: a multi-criteria assessment for Internet-of-Things (IoT) systems},
journal = {Enterprise Information Systems},
volume = {15},
number = {3},
pages = {332--351},
year = {2021},
publisher = {Taylor \& Francis},
doi = {10.1080/17517575.2019.1698060},


URL = { 
    
        https://doi.org/10.1080/17517575.2019.1698060
    
    

},
eprint = { 
    
        https://doi.org/10.1080/17517575.2019.1698060
    
    

}

}

@InProceedings{10.1007/978-3-030-44223-1_5,
author="Paquin, Christian
and Stebila, Douglas
and Tamvada, Goutam",
editor="Ding, Jintai
and Tillich, Jean-Pierre",
title="Benchmarking Post-quantum Cryptography in TLS",
booktitle="Post-Quantum Cryptography",
year="2020",
publisher="Springer International Publishing",
address="Cham",
pages="72--91",
isbn="978-3-030-44223-1"
}

@INPROCEEDINGS{11008381,
  author={Pajkos, Jaroslav and Kupcova, Eva and Pleva, Matus and Drutarovsky, Milos},
  booktitle={2025 35th International Conference Radioelektronika (RADIOELEKTRONIKA)}, 
  title={ESP32 Microcontroller based Lightweight TLS 1.3 Client for IoT Applications}, 
  year={2025},
  volume={},
  number={},
  pages={1-6},
  doi={10.1109/RADIOELEKTRONIKA65656.2025.11008381}}

@ARTICLE{10756206,
  author={Castiglione, Aniello and Esposito, Jacopo Gennaro and Loia, Vincenzo and Nappi, Michele and Pero, Chiara and Polsinelli, Matteo},
  journal={IEEE Transactions on Industrial Informatics}, 
  title={Integrating Post-Quantum Cryptography and Blockchain to Secure Low-Cost IoT Devices}, 
  year={2025},
  volume={21},
  number={2},
  pages={1674-1683},
  doi={10.1109/TII.2024.3485796}}

@INPROCEEDINGS{10821264,
  author={Sowa, Jakub and Hoang, Bach and Yeluru, Advaith and Qie, Steven and Nikolich, Anita and Iyer, Ravishankar and Cao, Phuong},
  booktitle={2024 IEEE International Conference on Quantum Computing and Engineering (QCE)}, 
  title={Post-Quantum Cryptography (PQC) Network Instrument: Measuring PQC Adoption Rates and Identifying Migration Pathways}, 
  year={2024},
  volume={01},
  number={},
  pages={1835-1846},
  doi={10.1109/QCE60285.2024.00213}}

@INPROCEEDINGS{8597401,
  author={Wukkadada, Bharati and Wankhede, Kirti and Nambiar, Ramith and Nair, Amala},
  booktitle={2018 International Conference on Inventive Research in Computing Applications (ICIRCA)}, 
  title={Comparison with HTTP and MQTT In Internet of Things (IoT)}, 
  year={2018},
  volume={},
  number={},
  pages={249-253},
  doi={10.1109/ICIRCA.2018.8597401}}

@ARTICLE{11197543,
  author={Zaheer, Ahmad Nawaz and Farhan, Muhammad and Naeem, Muhammad Rehan and Alnfiai, Mrim M.},
  journal={IEEE Transactions on Consumer Electronics}, 
  title={Quantum-Resilient Cryptographic Frameworks: Design and Analysis of Post-Quantum Algorithms for Secure and Efficient Edge-Assisted IoT Ecosystems in Consumer Electronics Devices}, 
  year={2025},
  volume={},
  number={},
  pages={1-1},
  doi={10.1109/TCE.2025.3619556}}

@article{Lezzi2025,
  author    = {Marianna Lezzi and Angelo Corallo and Mariangela Lazoi and Aldo Nimis},
  title     = {Measuring cyber resilience in industrial IoT: a systematic literature review},
  journal   = {Management Review Quarterly},
  year      = {2025},
  month     = apr,
  day       = {11},
  doi       = {10.1007/s11301-025-00495-8},
  url       = {https://doi.org/10.1007/s11301-025-00495-8},
  issn      = {2198-1639},
}

@inproceedings{10.1145/3587135.3592821,
author = {Tasopoulos, George and Dimopoulos, Charis and Fournaris, Apostolos P. and Zhao, Raymond K. and Sakzad, Amin and Steinfeld, Ron},
title = {Energy Consumption Evaluation of Post-Quantum TLS 1.3 for Resource-Constrained Embedded Devices},
year = {2023},
isbn = {9798400701405},
publisher = {Association for Computing Machinery},
address = {New York, NY, USA},
url = {https://doi.org/10.1145/3587135.3592821},
doi = {10.1145/3587135.3592821},
booktitle = {Proceedings of the 20th ACM International Conference on Computing Frontiers},
pages = {366–374},
numpages = {9},
location = {Bologna, Italy},
series = {CF '23}
}

@ARTICLE{11136103,
  author={Mrabet, Manel and Sliti, Maha},
  journal={IEEE Access}, 
  title={Toward Secure, Trustworthy, and Sustainable Edge Computing for Smart Cities: Innovative Strategies and Future Prospects}, 
  year={2025},
  volume={13},
  number={},
  pages={174236-174253},
  doi={10.1109/ACCESS.2025.3602390}}

@misc{halak2024securityassessmenttoolquantum,
      title={A Security Assessment tool for Quantum Threat Analysis}, 
      author={Basel Halak and Cristian Sebastian Csete and Edward Joyce and Jack Papaioannou and Alexandre Pires and Jin Soma and Betul Gokkaya and Michael Murphy},
      year={2024},
      eprint={2407.13523},
      archivePrefix={arXiv},
      primaryClass={cs.CR},
      url={https://arxiv.org/abs/2407.13523}, 
}

@article{PekPeterCh,
author = {Pekarcik, Peter and Chovancova, Eva},
year = {2025},
month = {09},
pages = {16-24},
title = {Post-Quantum Encryption Algorithms},
volume = {25},
journal = {Acta Electrotechnica et Informatica},
doi = {10.2478/aei-2025-0011}
}

@InProceedings{10.1007/978-981-32-9690-9_71,
author="Bhol, Seema Gupta
and Mohanty, J. R.
and Pattnaik, Prasant Kumar",
editor="Satapathy, Suresh Chandra
and Bhateja, Vikrant
and Mohanty, J. R.
and Udgata, Siba K.",
title="Cyber Security Metrics Evaluation Using Multi-criteria Decision-Making Approach",
booktitle="Smart Intelligent Computing and Applications ",
year="2020",
publisher="Springer Singapore",
address="Singapore",
pages="665--675",
isbn="978-981-32-9690-9"
}

@book{belton2002multiple,
  title        = {Multiple Criteria Decision Analysis: An Integrated Approach},
  author       = {Valerie Belton and Theodor J. Stewart},
  year         = {2002},
  publisher    = {Springer},
  address      = {New York, NY},
  edition      = {1},
  pages        = {XIX, 372},
  isbn         = {978-1-4615-1495-4},
  doi          = {10.1007/978-1-4615-1495-4},
  note         = {Originally published by Kluwer Academic Publishers, 2002; Hardcover ISBN: 978-0-7923-7505-0; Softcover ISBN: 978-1-4613-5582-3}
}

@article{Ameyed2023,
  author    = {Ameyed, Darine and Jaafar, Fehmi and Petrillo, Fabio and Cheriet, Mohamed},
  title     = {Quality and Security Frameworks for IoT-Architecture Models Evaluation},
  journal   = {SN Computer Science},
  year      = {2023},
  volume    = {4},
  number    = {4},
  pages     = {394},
  doi       = {10.1007/s42979-023-01815-z},
  url       = {https://doi.org/10.1007/s42979-023-01815-z}
}

@Article{cryptography9020032,
AUTHOR = {Abbasi, Maryam and Cardoso, Filipe and Váz, Paulo and Silva, José and Martins, Pedro},
TITLE = {A Practical Performance Benchmark of Post-Quantum Cryptography Across Heterogeneous Computing Environments},
JOURNAL = {Cryptography},
VOLUME = {9},
YEAR = {2025},
NUMBER = {2},
ARTICLE-NUMBER = {32},
URL = {https://www.mdpi.com/2410-387X/9/2/32},
ISSN = {2410-387X},
DOI = {10.3390/cryptography9020032}
}

@inproceedings{grgic2016web,
  title={A web-based IoT solution for monitoring data using MQTT protocol},
  author={Grgi{\'c}, Kre{\v{s}}imir and {\v{S}}peh, Ivan and Hejic, Ivan},
  booktitle={2016 International Conference on Smart Systems and Technologies (SST)},
  pages={249--253},
  year={2016},
  organization={IEEE}
}

@Article{iot6040062,
AUTHOR = {Jaddoa, Ali and Alharbi, Hasanein and Hommadi, Abbas and Ismael, Hussein A.},
TITLE = {Toward Scalable and Sustainable Detection Systems: A Behavioural Taxonomy and Utility-Based Framework for Security Detection in IoT and IIoT},
JOURNAL = {IoT},
VOLUME = {6},
YEAR = {2025},
NUMBER = {4},
ARTICLE-NUMBER = {62},
URL = {https://www.mdpi.com/2624-831X/6/4/62},
ISSN = {2624-831X},
DOI = {10.3390/iot6040062}
}

@Article{metrics2020009,
AUTHOR = {Soubra, Hassan and Elsayed, Hatem and Elbrolosy, Yousef and Adel, Youssef and Attia, Zeyad},
TITLE = {Comprehensive Review of Metrics and Measurements of Quantum Systems},
JOURNAL = {Metrics},
VOLUME = {2},
YEAR = {2025},
NUMBER = {2},
ARTICLE-NUMBER = {9},
URL = {https://www.mdpi.com/3042-5042/2/2/9},
ISSN = {3042-5042},
DOI = {10.3390/metrics2020009}
}

@book{boo1,
  author    = {Kulkarni, Anand},
  year      = {2022},
  month     = {02},
  pages     = {},
  title     = {Multiple Criteria Decision Making},
  isbn      = {978-981-16-7416-7},
  doi       = {10.1007/978-981-16-7414-3},
  publisher = {Springer}
}

@book{book,
author = {Nardo, Michela and Saisana, Michaela and Saltelli, Andrea and Tarantola, Stefano and Hoffman, Anders and Giovannini, Enrico},
year = {2008},
month = {09},
pages = {},
title = {Handbook on Constructing Composite Indicators and User Guide},
 publisher    = {Organisation for Economic Co‑operation and Development / Joint Research Centre of the European Commission},
volume = {2005},
isbn = {978-92-64-04345-9},
journal = {OECD Statistics Working Paper},
doi = {10.1787/533411815016}
}

@Article{smartcities8040116,
AUTHOR = {Rehman, Abdul and Alharbi, Omar},
TITLE = {QESIF: A Lightweight Quantum-Enhanced IoT Security Framework for Smart Cities},
JOURNAL = {Smart Cities},
VOLUME = {8},
YEAR = {2025},
NUMBER = {4},
ARTICLE-NUMBER = {116},
URL = {https://www.mdpi.com/2624-6511/8/4/116},
ISSN = {2624-6511},
DOI = {10.3390/smartcities8040116}
}

@inproceedings{10.1145/3716368.3735199,
author = {Dong, Ben and Wang, Qian},
title = {EPQUIC: Efficient Post-Quantum Cryptography for QUIC-Enabled Secure Communication},
year = {2025},
isbn = {9798400714962},
publisher = {Association for Computing Machinery},
address = {New York, NY, USA},
url = {https://doi.org/10.1145/3716368.3735199},
doi = {10.1145/3716368.3735199},
booktitle = {Proceedings of the Great Lakes Symposium on VLSI 2025},
pages = {141–146},
numpages = {6},
location = {
},
series = {GLSVLSI '25}
}

@Article{electronics14214234,
AUTHOR = {Iliadis-Apostolidis, Dimosthenis and Lawo, Daniel Christian and Kosta, Sokol and Tafur Monroy, Idelfonso and  Vegas Olmos, Juan Jose},
TITLE = {QRoNS: Quantum Resilience over IPsec Tunnels for Network Slicing},
JOURNAL = {Electronics},
VOLUME = {14},
YEAR = {2025},
NUMBER = {21},
ARTICLE-NUMBER = {4234},
URL = {https://www.mdpi.com/2079-9292/14/21/4234},
ISSN = {2079-9292},
DOI = {10.3390/electronics14214234}
}

@article{jara2023comparative,
  title={Comparative analysis of power consumption between MQTT and HTTP protocols in an IoT platform designed and implemented for remote real-time monitoring of long-term cold chain transport operations},
  author={Jara Ochoa, Heriberto J and Pe{\~n}a, Raul and Ledo Mezquita, Yoel and Gonzalez, Enrique and Camacho-Leon, Sergio},
  journal={Sensors},
  volume={23},
  number={10},
  pages={4896},
  year={2023},
  publisher={MDPI}
}

@article{gentile2024network,
  title={A Network Performance Analysis of MQTT Security Protocols with Constrained Hardware in the Dark Net for DMS.},
  author={Gentile, Antonio Francesco and Macr{\`\i}, Davide and Carn{\`\i}, Domenico Luca and Greco, Emilio and Lamonaca, Francesco},
  journal={Applied Sciences (2076-3417)},
  volume={14},
  number={18},
  year={2024}
}

@article{ghotbou2021comparing,
  title={Comparing application layer protocols for video transmission in IoT low power lossy networks: An analytic comparison},
  author={Ghotbou, Arvin and Khansari, Mohammad},
  journal={Wireless Networks},
  volume={27},
  number={1},
  pages={269--283},
  year={2021},
  publisher={Springer}
}

@article{sadowski2018rssi,
  title={Rssi-based indoor localization with the internet of things},
  author={Sadowski, Sebastian and Spachos, Petros},
  journal={IEEE access},
  volume={6},
  pages={30149--30161},
  year={2018},
  publisher={IEEE}
}

@inproceedings{hernandez2020lightweight,
  title={Lightweight and standalone IoT based WiFi sensing for active repositioning and mobility},
  author={Hernandez, Steven M and Bulut, Eyuphan},
  booktitle={2020 IEEE 21st International Symposium on" A World of Wireless, Mobile and Multimedia Networks"(WoWMoM)},
  pages={277--286},
  year={2020},
  organization={IEEE}
}

@INPROCEEDINGS{10763879,
  author={Chakraborty, Amit},
  booktitle={2024 IEEE International Conference on Communication, Computing and Signal Processing (IICCCS)}, 
  title={Multi-Criteria Decision Analysis Framework for Optimal Combination of Blockchain to Determine Scalability in IoT}, 
  year={2024},
  volume={},
  number={},
  pages={1-6},
  doi={10.1109/IICCCS61609.2024.10763879}}

@inproceedings{rampazzo2023assessment,
  title={Assessment of the Impact of Hybrid Post-Quantum Cryptography on the Performance of the MQTT Communication Protocol},
  author={Rampazzo, Felipe Jos{\'e} Aguiar and Henriques, Marco Aur{\'e}lio Amaral},
  booktitle={2023 Symposium on Internet of Things (SIoT)},
  pages={1--5},
  year={2023},
  organization={IEEE}
}

@article{asif2021post,
  title={Post-quantum cryptosystems for Internet-of-Things: A survey on lattice-based algorithms},
  author={Asif, Rameez},
  journal={IoT},
  volume={2},
  number={1},
  pages={71--91},
  year={2021},
  publisher={Multidisciplinary Digital Publishing Institute}
}

@inproceedings{abdelhafeez2019study,
  title={A Study on Transmission Overhead of Post Quantum Cryptography Algorithms in Internet of Things Networks},
  author={AbdelHafeez, Mahmoud and Taha, Mostafa and Khaled, Elsayed Esam M and AbdelRaheem, Mohamed},
  booktitle={2019 31st International Conference on Microelectronics (ICM)},
  pages={113--117},
  year={2019},
  organization={IEEE}
}

@article{detti2020sub,
  title={Sub-linear scalability of MQTT clusters in topic-based publish-subscribe applications},
  author={Detti, Andrea and Funari, Ludovico and Blefari-Melazzi, Nicola},
  journal={IEEE Transactions on Network and Service Management},
  volume={17},
  number={3},
  pages={1954--1968},
  year={2020},
  publisher={IEEE}
}

@article{hong2023performance,
  title={Performance comparison of http, https, and mqtt for iot applications},
  author={Hong, Sukjun and Kang, Jinkyu and Kwon, Soonchul},
  journal={The International Journal of Advanced Smart Convergence},
  volume={12},
  number={1},
  pages={9--17},
  year={2023}
}

@article{akpakwu2025congestion,
  title={Congestion Control in Constrained Application Protocol for the Internet of Things: State-of-the-Art, Challenges and Future Directions},
  author={Akpakwu, Godfrey A and Mathonsi, Topside E and Tshilongamulenze, Tshimangadzo M and Maswikaneng, Solly P and Muchenje, Tonderai},
  journal={IEEE Access},
  year={2025},
  publisher={IEEE}
}

@incollection{leggieri2013interoperability,
  title={Interoperability of two RESTful protocols: HTTP and CoAP},
  author={Leggieri, Myriam and Hausenblas, Michael},
  booktitle={Rest: Advanced research topics and practical applications},
  pages={27--49},
  year={2013},
  publisher={Springer}
}

@article{scott2023tls,
  title={On TLS for the Internet of Things, in a Post Quantum world},
  author={Scott, Michael},
  journal={Cryptology ePrint Archive},
  year={2023}
}

@inproceedings{nofal2019comprehensive,
  title={A comprehensive empirical analysis of tls handshake and record layer on iot platforms},
  author={Nofal, Ramzi A and Tran, Nam and Garcia, Carlos and Liu, Yuhong and Dezfouli, Behnam},
  booktitle={Proceedings of the 22nd International ACM Conference on Modeling, Analysis and Simulation of Wireless and Mobile Systems},
  pages={61--70},
  year={2019}
}

@article{yang2013rssi,
  title={From RSSI to CSI: Indoor localization via channel response},
  author={Yang, Zheng and Zhou, Zimu and Liu, Yunhao},
  journal={ACM Computing Surveys (CSUR)},
  volume={46},
  number={2},
  pages={1--32},
  year={2013},
  publisher={ACM New York, NY, USA}
}

@inproceedings{el2018performance,
  title={Performance analysis of received signal strength and link quality in wireless sensor networks},
  author={El Houssaini, Dhouha and Khriji, Sabrine and Besbes, Kamel and Kanoun, Olfa},
  booktitle={2018 15th International Multi-Conference on Systems, Signals \& Devices (SSD)},
  pages={173--178},
  year={2018},
  organization={IEEE}
}

@inproceedings{lui2011differences,
  title={Differences in RSSI readings made by different Wi-Fi chipsets: A limitation of WLAN localization},
  author={Lui, Gough and Gallagher, Thomas and Li, Binghao and Dempster, Andrew G and Rizos, Chris},
  booktitle={2011 International conference on localization and GNSS (ICL-GNSS)},
  pages={53--57},
  year={2011},
  organization={IEEE}
}

@article{pereira2017performance,
  title={Performance evaluation of cryptographic algorithms over IoT platforms and operating systems},
  author={Pereira, Geovandro CCF and Alves, Renan CA and Silva, Felipe L da and Azevedo, Roberto M and Albertini, Bruno C and Margi, C{\'\i}ntia B},
  journal={Security and Communication Networks},
  volume={2017},
  number={1},
  pages={2046735},
  year={2017},
  publisher={Wiley Online Library}
}

@INPROCEEDINGS{7498318,
  author={Franzke, Maximilian and Emrich, Tobias and Züfle, Andreas and Renz, Matthias},
  booktitle={2016 IEEE 32nd International Conference on Data Engineering (ICDE)}, 
  title={Indexing multi-metric data}, 
  year={2016},
  volume={},
  number={},
  pages={1122-1133},
  doi={10.1109/ICDE.2016.7498318}}

@inproceedings{sobral2016composite,
  title={A composite routing metric for wireless sensor networks in AAL-IoT},
  author={Sobral, Jos{\'e} VV and Rodrigues, Joel JPC and Saleem, Kashif and De Paz, Juan F and Corchado, Juan M},
  booktitle={2016 9th IFIP Wireless and Mobile Networking Conference (WMNC)},
  pages={168--173},
  year={2016},
  organization={IEEE}
}

@article{krishnamurthi2020overview,
  title={An overview of IoT sensor data processing, fusion, and analysis techniques},
  author={Krishnamurthi, Rajalakshmi and Kumar, Adarsh and Gopinathan, Dhanalekshmi and Nayyar, Anand and Qureshi, Basit},
  journal={Sensors},
  volume={20},
  number={21},
  pages={6076},
  year={2020},
  publisher={MDPI}
}

@article{al2020survey,
  title={A survey of machine and deep learning methods for internet of things (IoT) security},
  author={Al-Garadi, Mohammed Ali and Mohamed, Amr and Al-Ali, Abdulla Khalid and Du, Xiaojiang and Ali, Ihsan and Guizani, Mohsen},
  journal={IEEE communications surveys \& tutorials},
  volume={22},
  number={3},
  pages={1646--1685},
  year={2020},
  publisher={IEEE}
}

@article{hussain2020machine,
  title={Machine learning in IoT security: Current solutions and future challenges},
  author={Hussain, Fatima and Hussain, Rasheed and Hassan, Syed Ali and Hossain, Ekram},
  journal={IEEE Communications Surveys \& Tutorials},
  volume={22},
  number={3},
  pages={1686--1721},
  year={2020},
  publisher={IEEE}
}

@article{kataria2022ai,
  title={AI-and IoT-based hybrid model for air quality prediction in a smart city with network assistance},
  author={Kataria, Aman and Puri, Vikram},
  journal={IET networks},
  volume={11},
  number={6},
  pages={221--233},
  year={2022},
  publisher={Wiley Online Library}
}

@article{gencay1997nonlinear,
  title={Nonlinear modelling and prediction with feedforward and recurrent networks},
  author={Gencay, Ramazan and Liu, Tung},
  journal={Physica D: Nonlinear Phenomena},
  volume={108},
  number={1-2},
  pages={119--134},
  year={1997},
  publisher={Elsevier}
}

@article{darzi2024pqc,
  title={Pqc meets ml or ai: Exploring the synergy of machine learning and post-quantum cryptography},
  author={Darzi, Saleh and Yavuz, Attila A},
  journal={Authorea Preprints},
  year={2024},
  publisher={Authorea}
}

@article{huo2021performance,
  title={Performance prediction of proton-exchange membrane fuel cell based on convolutional neural network and random forest feature selection},
  author={Huo, Weiwei and Li, Weier and Zhang, Zehui and Sun, Chao and Zhou, Feikun and Gong, Guoqing},
  journal={Energy Conversion and Management},
  volume={243},
  pages={114367},
  year={2021},
  publisher={Elsevier}
}

@article{tekin2023energy,
  title={Energy consumption of on-device machine learning models for IoT intrusion detection},
  author={Tekin, Nazli and Acar, Abbas and Aris, Ahmet and Uluagac, A Selcuk and Gungor, Vehbi Cagri},
  journal={Internet of Things},
  volume={21},
  pages={100670},
  year={2023},
  publisher={Elsevier}
}

@inproceedings{de2018identifying,
  title={Identifying and detecting applications within TLS traffic},
  author={De Lucia, Michael J and Cotton, Chase},
  booktitle={Cyber Sensing 2018},
  volume={10630},
  pages={179--190},
  year={2018},
  organization={SPIE}
}

@inproceedings{dowling2015cryptographic,
  title={A cryptographic analysis of the TLS 1.3 handshake protocol candidates},
  author={Dowling, Benjamin and Fischlin, Marc and G{\"u}nther, Felix and Stebila, Douglas},
  booktitle={Proceedings of the 22nd ACM SIGSAC conference on computer and communications security},
  pages={1197--1210},
  year={2015}
}

@article{ayan2023comprehensive,
  title={A comprehensive review of the novel weighting methods for multi-criteria decision-making},
  author={Ayan, B{\"u}{\c{s}}ra and Abac{\i}o{\u{g}}lu, Seda and Basilio, Marcio Pereira},
  journal={Information},
  volume={14},
  number={5},
  pages={285},
  year={2023},
  publisher={MDPI}
}

@incollection{shafik2024machine,
  title={Machine learning techniques for multicriteria decision-making},
  author={Shafik, Wasswa},
  booktitle={Multi-Criteria Decision-Making and Optimum Design with Machine Learning},
  pages={165--194},
  year={2024},
  publisher={CRC Press}
}

@article{wang2025review,
  title={A Review on the Advances, Applications, and Future Prospects of Post-Quantum Cryptography in Blockchain, IoT},
  author={Wang, Yong and Ismail, Eddie Shahril},
  journal={IEEE Access},
  year={2025},
  publisher={IEEE}
}

\end{document}